\documentclass[prd,nofootinbib,showkeys,showpacs]{revtex4}
\usepackage{amsmath}

\makeatletter
\newcommand{\contraction}[5][1ex]{%
  \mathchoice
    {\contraction@\displaystyle{#2}{#3}{#4}{#5}{#1}}%
    {\contraction@\textstyle{#2}{#3}{#4}{#5}{#1}}%
    {\contraction@\scriptstyle{#2}{#3}{#4}{#5}{#1}}%
    {\contraction@\scriptscriptstyle{#2}{#3}{#4}{#5}{#1}}}%
\newcommand{\contraction@}[6]{%
  \setbox0=\hbox{$#1#2$}%
  \setbox2=\hbox{$#1#3$}%
  \setbox4=\hbox{$#1#4$}%
  \setbox6=\hbox{$#1#5$}%
  \dimen0=\wd2%
  \advance\dimen0 by \wd6%
  \divide\dimen0 by 2%
  \advance\dimen0 by \wd4%
  \vbox{%
    \hbox to 0pt{%
      \kern \wd0%
      \kern 0.5\wd2%
      \contraction@@{\dimen0}{#6}%
      \hss}%
    \vskip 0.5ex
    \vskip\ht2}}

\newcommand{\contraction@@}[3][0.05em]{%
  \hbox{%
    \vrule width #1 height 0pt depth #3%
    \vrule width #2 height 0pt depth #1%
    \vrule width #1 height 0pt depth #3%
    \relax}}
\makeatother

\begin{document}

\title{Factorization and Non-Factorization of In-Medium Four-Quark Condensates}

\author{Stefan Leupold}


\affiliation{Institut f\"ur Theoretische Physik, Universit\"at Giessen, Germany}

\begin{abstract}
It is well-established for the vacuum case that in the limit of a large number of colors
$N_c$ the four-quark 
condensates factorize into products of the two-quark condensate. 
It is shown that in the combined large-$N_c$ and linear-density approximation
four-quark condensates do not factorize in a medium of pions (finite temperature system)
but do factorize in a medium of nucleons (nuclear system).
\end{abstract}

\maketitle

It is by now well accepted that QCD, the theory of the strong interaction, has a 
non-trivial vacuum structure. One way to characterize this structure is by means of
non-vanishing matrix elements of quark or gluon operators, the condensates. Particular
condensates also play an important role for the connection of hadronic properties
to the underlying quark-gluon structure, as investigated in much detail within the QCD
sum rule method \cite{shif79,Reinders:1984sr}. Most prominently, the two-quark condensate,
the gluon condensate and four-quark condensates appear in the sum rules. 
When it comes to numbers
the largest uncertainties reside in the actual values for the four-quark condensates.
They remained a matter of constant debate over the last decades 
\cite{shif79,Launer:1983ib,Bertlmann:1987ty,%
Dominguez:1987nw,Gimenez:1990vg,Leinweber:1995fn,klingl2,Leupold:1998dg}. 
One central point in these
discussions is the question whether a four-quark condensate can be factorized more or
less accurately into a product of two-quark condensates.

In hot and/or dense enough strongly 
interacting media, QCD is subject to phase transitions or rapid 
crossovers \cite{Karsch:2001cy,Rajagopal:2000wf}. Consequently,
the condensates typically change with density and temperature. In lowest non-trivial 
order in temperature and density the changes of the two-quark and the gluon
condensate are well established \cite{Gerber:1989tt,Drukarev:1991fs,%
Hatsuda:1993bv,Hatsuda:1992ez}. 
The issue is much less clear for the four-quark 
condensates \cite{Eletsky:1992xd,Hatsuda:1993bv,klingl2,Leupold:1998dg,%
Zschocke:2002mp,Zschocke:2003hx}.
To summarize, for the vacuum values as well as for the in-medium changes the uncertainties
are the largest for the four-quark condensates. 

There is one particular limit of QCD, however, where exact relations can be obtained, 
namely the limit of a large number of colors $N_c$. While the
physical ($N_c = 3$) values of four-quark condensates are a matter of debate, at least
in the large-$N_c$ limit the factorization into two-quark condensates holds for the 
vacuum case \cite{Novikov:1984jt}. In the present work we generalize this aspect to
typical in-medium cases. We will show that in the large-$N_c$ limit 
an arbitrary four-quark condensate factorizes
in a nuclear medium, while it does not in a pionic medium. The pion case has already been
discussed in detail in \cite{Eletsky:1992xd,Hatsuda:1993bv} from a 
different point of view. The nucleon case was addressed only recently by the 
present author \cite{Leupold:2004gh}. In the following, we
present a unified derivation for pions and nucleons.

We start by recalling how some hadronic properties scale 
as a function of the number of colors
$N_c$. According to \cite{'tHooft:1974jz,witten} the following rules hold for the
expansion of hadronic quantities in powers of $1/N_c$:
\begin{itemize}
\item Mesons are $q \bar q$-states, their masses are $O(N_c^0)$.
\item In contrast, a baryon consists of $N_c$ quarks, its mass is $O(N_c)$.
\item The ratio between a quark current $A$ and the 
corresponding interpolating hadronic field $A_H$ is $O(\sqrt{N_c})$.
\item Mesonic interactions are suppressed. Generically $n$-meson vertices are only 
$O(N_c^{1-n/2})$. Consequently, the decay width
of a meson into two other mesons is $O(1/N_c)$. Therefore, in vacuum mesons are
stable in the large-$N_c$ limit.
\item In contrast, meson-baryon interactions are not suppressed. Meson-baryon 
scattering amplitudes are $O(N_c^0)$. 
A meson-baryon-baryon vertex can even be enhanced as
$O(\sqrt{N_c})$. (Subtle cancellations ensure that these two rules are compatible, 
cf.~e.g.~\cite{Lam:1997ry}.)
\end{itemize}

Using Fierz transformations, in particular \cite{pastar} 
\begin{equation}
  \label{eq:pastar}
{1 \over 2} (\lambda_a)_{\alpha\beta} \, (\lambda_a)_{\gamma\delta} = 
\delta_{\alpha\delta} \delta_{\beta\gamma} 
- {1 \over N_c} \, \delta_{\alpha\beta} \delta_{\gamma\delta} \,,
\end{equation}
every four-quark operator 
\begin{equation}
  \label{eq:four-lambda}
\bar\psi \lambda_a \Gamma \psi \, \bar \psi \lambda_a \Gamma' \psi
\end{equation}
can be written as a sum of
products of two color neutral two-quark operators 
\begin{equation}
  \label{eq:two-qu}
\bar \psi \Gamma'' \psi \,. 
\end{equation}
Here $\lambda_a$ denotes a color matrix and $\Gamma$, $\Gamma'$, $\Gamma''$ 
matrices in spinor and flavor space. Therefore we can restrict our
considerations to four-quark condensates of type $\langle A B \rangle_{\rm med.}$ 
where $A$ and $B$ are color neutral two-quark (to be precise quark-antiquark) operators
of type \eqref{eq:two-qu}. In the following we will interpret $A$ and $B$ as interpolating
fields for hadrons. Effectively we will assume that color is saturated within $A$ and $B$
separately. In other words: The quark and the antiquark field of $A$ form a hadron, same for
$B$. In that way, however, we neglect possible exchange terms, i.e.~the ones where 
e.g.~the quark field from $A$ and the antiquark field from $B$ form a hadron. Fortunately
these exchange terms are subleading in the number of colors and therefore of no concern 
for our present considerations. This can be seen as follows:
\begin{equation}
  \label{eq:direct}
\contraction{}{\bar q_i}{\Gamma}{q_i}\bar q_i \Gamma q_i \, 
\contraction{}{\bar q_j}{\Gamma'}{q_j}\bar q_j \Gamma' q_j  
\to \delta_{ii} \delta_{jj} = N_c^2
\end{equation}
whereas
\begin{equation}
  \label{eq:exchange}
\contraction{\bar q_i \Gamma}{q_i}{\,}{\bar q_j}
\contraction[1.5ex]{}{\bar q_i}{\Gamma q_i \, \bar q_j \Gamma'}{q_j}
\bar q_i \Gamma q_i \, \bar q_j \Gamma' q_j
\to \delta_{ij} \delta_{ij} = N_c \,.
\end{equation}
Here the connection $\contraction{}{a}{i}{b}\ldots$ 
indicates that the two respective fields form hadronic states.

We approximate the medium by a gas of non-interacting states of type $X$ with
density $\rho_X$. This is a valid approximation as long as the density is not too
high. The extension to a gas composed of several species is straightforward.
For a system at finite temperature (nuclear density) one uses pions (nucleons)
for $X$. We get
\begin{equation}
  \label{eq:app-fourmed}
\langle A B \rangle_{\rm med.} \approx \langle 0 \vert A B \vert 0 \rangle
+ \rho_X \langle X \vert A B \vert X \rangle
\end{equation}
where $\vert 0 \rangle$ denotes the vacuum state.
The expectation value with respect to the single state $X$ consists of five parts: 
First, the transition process 
$A+X$ to $X$, with $B$ as a pure spectator; second, the same process with
the roles of $A$ and $B$ reversed; third, the annihilation of $X$ by $B$ and the
creation of $X$ by $A$; fourth, the latter process with the roles of $A$ and $B$ reversed; 
finally, the true scattering of 
$B$ with $X$ into $A$ and $X$. Denoting the true scattering process by 
$\langle X \vert AB \vert X \rangle_{\rm connected}$ we get the following decomposition:
\begin{eqnarray}
\langle X \vert AB \vert X \rangle & = &
\langle X \vert A \vert X \rangle \, \langle 0 \vert B \vert 0 \rangle
+ \langle 0 \vert A \vert 0 \rangle \, \langle X \vert B \vert X \rangle
+ \langle X \vert A \vert 0 \rangle \, \langle 0 \vert B \vert X \rangle 
+ \langle 0 \vert A \vert X \rangle \, \langle X \vert B \vert 0 \rangle 
+\langle X \vert AB \vert X \rangle_{\rm connected}  \,. \phantom{mm}
  \label{eq:app-nuclexp}
\end{eqnarray}

So far, everything was rather general. In particular, we have not involved any 
large-$N_c$ arguments besides the dropping of the exchange terms. 
Now we recall that a quark current $A$ is connected
to mesonic fields $A_H$ with the same quantum numbers by a factor which is 
$O(\sqrt{N_c})$. Hence the connected part 
in (\ref{eq:app-nuclexp}) is of order $N_c$
times a hadronic scattering amplitude. The size of the latter depends on $X$, but
it is at most $O(N_c^0)$. Below we will find that this term is always subleading.
Independent of $X$ we can already factorize the vacuum part
in (\ref{eq:app-fourmed}) in the large-$N_c$ approximation:
\begin{eqnarray}
\langle A B \rangle_{\rm med.} & \approx &
\langle 0 \vert A \vert 0 \rangle \, \langle 0 \vert B \vert 0 \rangle + o(N_c) 
\nonumber \\ && {}
+ \rho_X \langle X \vert A \vert X \rangle \, \langle 0 \vert B \vert 0 \rangle
+ \rho_X \langle 0 \vert A \vert 0 \rangle \, \langle X \vert B \vert X \rangle
\nonumber \\ && {}
+ \rho_X \langle X \vert A \vert 0 \rangle \, \langle 0 \vert B \vert X \rangle
+ \rho_X \langle 0 \vert A \vert X \rangle \, \langle X \vert B \vert 0 \rangle
+ \rho_X \langle X \vert AB \vert X \rangle_{\rm connected}  \,.
  \label{eq:app-fourmed2}
\end{eqnarray}
Note that the leading part here is $O(N_c^2)$ since 
$\langle 0 \vert A \vert 0 \rangle = O(N_c)$. 

On the right hand side of \eqref{eq:app-fourmed2} the terms given explicitly in the
first two lines constitute the result of a factorization assumption:
\begin{eqnarray}
\lefteqn{\langle A \rangle_{\rm med.} \, \langle B \rangle_{\rm med.} =
\left( 
\langle 0 \vert A \vert 0 \rangle + \rho_X \langle X \vert A \vert X \rangle 
\right) \,
\left( 
\langle 0 \vert B \vert 0 \rangle + \rho_X \langle X \vert B \vert X \rangle 
\right) 
=} \nonumber \\ &&
\langle 0 \vert A \vert 0 \rangle \, \langle 0 \vert B \vert 0 \rangle 
+ \rho_X \langle X \vert A \vert X \rangle \, \langle 0 \vert B \vert 0 \rangle
+ \rho_X \langle 0 \vert A \vert 0 \rangle \, \langle X \vert B \vert X \rangle
+ o(\rho_X^2)   \,.
  \label{eq:factass}
\end{eqnarray}
(Note that we disregard terms quadratic in the density as we work in the linear-density
approximation.) On the other hand, 
the terms given in the last line of \eqref{eq:app-fourmed2} in 
general spoil the factorization assumption. In the following we will find that in leading
order of a $1/N_c$ expansion these terms are present for the case of finite temperatures
but absent for the case of finite baryon densities.

To get a first glance how the large-$N_c$ rules work we shall prove the vacuum 
factorization: We start by inserting a (hadronic) unity operator between $A$ and $B$ 
and get
\begin{equation}
  \label{eq:vacfac}
\langle 0 \vert A B \vert 0 \rangle = 
\langle 0 \vert A \vert 0 \rangle \, \langle 0 \vert B \vert 0 \rangle
+ \sum\limits_h \, \langle 0 \vert A \vert h \rangle \, \langle h \vert B \vert 0 \rangle
+ \mbox{insertions of multi-particle states.}
\end{equation}
As already stated the first term on the right hand side is $O(N_c^2)$. The next term is
only $O(N_c)$ due to
\begin{equation}
  \label{eq:onehadr}
\langle 0 \vert A \vert h \rangle \sim \sqrt{N_c} \, \langle 0 \vert A_H \vert h \rangle
\sim \sqrt{N_c}  \,.
\end{equation}
The multi-particle states are even further suppressed. Hence in the vacuum any four-quark
condensate factorizes in leading order of the $1/N_c$ expansion.

To make further progress we have to specify $X$. On account of the rules discussed above
we have for pions
\begin{equation}
  \label{eq:app-pion-two}
\langle \pi \vert A \vert \pi \rangle \sim  
\sqrt{N_c} \, \langle \pi \vert A_H \vert \pi \rangle \sim N_c^0  \,,
\end{equation}
\begin{equation}
  \label{eq:app-pion-vac}
\langle \pi \vert A \vert 0 \rangle \sim 
\sqrt{N_c} \, \langle \pi \vert A_H \vert 0 \rangle \sim \sqrt{N_c} 
\end{equation}
and
\begin{equation}
  \label{eq:app-pion-conn}
\langle \pi \vert A B \vert \pi \rangle_{\rm connected} \sim
N_c \, \langle \pi \vert A_H B_H \vert \pi \rangle_{\rm connected} \sim N_c^0  \,,
\end{equation}
whereas for nucleons we find
\begin{equation}
  \label{eq:app-nucl-two}
\langle N \vert A \vert N \rangle \sim  
\sqrt{N_c} \, \langle N \vert A_H \vert N \rangle\sim N_c  \,,
\end{equation}
\begin{equation}
  \label{eq:app-nucl-vac}
\langle N \vert A \vert 0 \rangle = 0 \,,
\end{equation}
and
\begin{equation}
  \label{eq:app-nucl-conn}
\langle N \vert A B \vert N \rangle_{\rm connected} \sim
N_c \, \langle N \vert A_H B_H \vert N \rangle_{\rm connected} \sim N_c  \,.
\end{equation}
Of course, all relations displayed for $A$ hold also for $B$.
Note that $A$ and $B$ are quark-antiquark operators. Hence they cannot create
a nucleon out of the vacuum. Therefore $\langle N \vert A \vert 0 \rangle$ has to
vanish as stated in (\ref{eq:app-nucl-vac}). This will constitute the important difference
which makes the temperature and the nuclear density case distinct from each other.

We conclude, first of all, that all pionic in-medium effects are suppressed as 
compared to the vacuum part:
\begin{eqnarray}
\langle A B \rangle_{\rm pionic \; med.} & \approx &
\langle 0 \vert A \vert 0 \rangle \, \langle 0 \vert B \vert 0 \rangle + o(N_c) \,.
  \label{eq:app-fourmed3}
\end{eqnarray}
If we keep the highest non-trivial order in $N_c$ of the medium part, we get
\begin{eqnarray}
\lefteqn{
\langle A B \rangle_{\rm pionic \; med.} \approx 
\langle 0 \vert A \vert 0 \rangle \, \langle 0 \vert B \vert 0 \rangle + o(N_c)} 
\nonumber \\ && {}
+ \rho_\pi \left(
\langle \pi \vert A \vert \pi \rangle \, \langle 0 \vert B \vert 0 \rangle
+ \langle 0 \vert A \vert 0 \rangle \, \langle \pi \vert B \vert \pi \rangle
+ \langle \pi \vert A \vert 0 \rangle \, \langle 0 \vert B \vert \pi \rangle
+ \langle 0 \vert A \vert \pi \rangle \, \langle \pi \vert B \vert 0 \rangle
+ o(N_c^0)
\right)  \,. \phantom{mmm}
  \label{eq:app-fourmed3a}
\end{eqnarray}
Obviously even now factorization is spoiled by the annihilation and creation terms like 
$\langle \pi \vert A \vert 0 \rangle \, \langle 0 \vert B \vert \pi \rangle$.
Such terms do not vanish, if $A$ and $B$ have the quantum numbers of the 
pion.\footnote{Note that the matrix elements which appear in \eqref{eq:app-fourmed3a}
can be calculated using current algebra \cite{Eletsky:1992xd,Hatsuda:1993bv}.}

In contrast, in a nuclear medium we find ``factorization'':
\begin{eqnarray}
\langle A B \rangle_{\rm nuclear \; med.} & \approx &
\langle 0 \vert A \vert 0 \rangle \, \langle 0 \vert B \vert 0 \rangle 
+ \rho_N \langle N \vert A \vert N \rangle \, \langle 0 \vert B \vert 0 \rangle
+ \rho_N \langle 0 \vert A \vert 0 \rangle \, \langle N \vert B \vert N \rangle 
+ o(N_c)  \,.
  \label{eq:app-fourmed4}
\end{eqnarray}
The quotation marks are meant to indicate that there is no $\rho_N^2$ term which
would appear if factorization was taken literally.

To summarize, we have addressed a limit in which for the vacuum case
the four-quark condensates can be expressed in terms of the two-quark condensate, namely
the large-$N_c$ limit. We have shown that in the same limit we can make exact statements
for in-medium expectation values of four-quark condensates. In particular, reproducing
the results of \cite{Eletsky:1992xd,Hatsuda:1993bv} with a different technique 
we have found
that in the commonly used linear-density approximation in general four-quark condensates 
do not factorize for the finite-temperature case. 
In addition, we have found that (also in linear-density approximation) for
the nuclear medium four-quark condensates factorize into two-quark condensates. 
This qualitative difference has a simple reason: Among the
terms which potentially spoil the factorization (last line in \eqref{eq:app-fourmed2}) 
the true scattering process is always subleading in $1/N_c$ as compared to the in-medium
terms which show up in a factorization assumption (cf.~\eqref{eq:factass}). 
The creation and annihilation terms, on the other hand, --- if present ---
contribute with the same power in $1/N_c$ as the terms in \eqref{eq:factass}. In a mesonic
medium these annihilation and creation terms can be non-vanishing whereas in a baryonic
medium they vanish. 

Finally, we
note that with the decomposition \eqref{eq:app-nuclexp} and the interpretation of
$\langle X \vert AB \vert X \rangle_{\rm connected}$ as a true scattering amplitude
it might be feasible to estimate this amplitude from hadronic models. (This is similar
in spirit to the approach presented in \cite{Eletsky:1992rs}.) Especially the experimentally
well explored pion-nucleon scattering amplitude can be used. First results in that direction
are rather satisfying as they indicate that corrections to \eqref{eq:app-fourmed4}
are quite small. A more detailed investigation is now in progress. 
Besides phenomenology again the
large-$N_c$ expansion might be helpful: For a given hadronic model (e.g.~with the
well-known physical states $\pi$, $\rho$, $\omega$, $N$ and $\Delta$) one can calculate
meson-nucleon scattering in the large-$N_c$ limit to all loop 
orders \cite{Arnold:1990fp,Mattis:1994ra}.
This would yield the next-to-leading $O(N_c)$ term in \eqref{eq:app-fourmed4}.
In that
way the validity of the in-medium factorization \eqref{eq:app-fourmed4} could be checked
beyond the leading order of the $1/N_c$ expansion.

\acknowledgments The author thanks A.~Peshier for discussions and for reading 
the manuscript and U.~Mosel for continuous support.

\bibliography{literature}
\bibliographystyle{apsrev}

\end{document}